\documentstyle[aaspp]{article}

\newcommand{\bq}{\begin{equation}}
\newcommand{\eq}{\end{equation}}

\begin{document}
\def\refitem{\par\parskip 0pt\noindent\hangindent 20pt}
 
\title{Tests of the Accelerating Universe \\ with Near-Infrared Observations of a High-Redshift Type Ia Supernova}

Adam G. Riess\footnote{Space Telescope Science Institute, 3700 San Martin Drive, 
Baltimore, MD 21218},  Alexei V. Filippenko\footnote{Department of Astronomy, University of California, 
Berkeley, CA 94720-3411}, Michael C. Liu$^{2}$, 
Peter Challis\footnote{Harvard-Smithsonian Center for Astrophysics, 60
Garden St., Cambridge, MA 02138},
 Alejandro Clocchiatti\footnote{Departamento de Astronom\'{\i}a y
 Astrof\'{\i}sica
Pontificia Universidad Cat\'olica, Casilla 104, Santiago 22, Chile},
Alan Diercks\footnote{Department of Astronomy, University of Washington, Seattle, WA 
98195},
Peter M. Garnavich$^3$, Craig J. Hogan$^5$,
Saurabh Jha$^3$, Robert P. Kirshner$^3$,
B. Leibundgut\footnote{European Southern Observatory,
 Karl-Schwarzschild-Strasse 2,  Garching, Germany},
M. M. Phillips\footnote{Cerro Tololo Inter-American Observatory,
Casilla 603, La Serena, Chile.  NOAO is operated by the Association of Universities for Research in Astronomy (AURA) under
cooperative agreement with the National Science Foundation.},
David Reiss$^{5}$, Brian P. Schmidt\footnote{Mount Stromlo and Siding Spring Observatories,
Private Bag, Weston Creek P.O. 2611,  Australia}
\footnote{Visiting astronomer, Cerro Tololo Inter-American
Observatory, National Optical Astronomy Observatories, operated by the
Association of Universities for Research in Astronomy (AURA) under
cooperative agreement with the National Science Foundation.},
 Robert A. Schommer$^7$,
R. Chris Smith$^{7\,}$\footnote{University of Michigan, Department of Astronomy,
 834 
Dennison Bldg., Ann Arbor, MI 48109},
J. Spyromilio$^{6}$,
Christopher Stubbs$^{5}$,
Nicholas B. Suntzeff$^7$,
John Tonry\footnote{Institute for Astronomy, University of Hawaii, 2680
Woodlawn Dr., Honolulu, HI 96822}, Patrick Woudt$^6$, Robert J. Brunner\footnote{Department of Astronomy, 105-24 Caltech, Pasadena, CA 91125}, Arjun Dey\footnote{National Optical Astronomy Observatories, Tucson, AZ 85719}, Roy Gal$^{12}$, James Graham$^2$, James Larkin\footnote{Department of Physics \& Astronomy, University of California, Los Angelos, CA 90095-1562}, Steve C. Odewahn$^{12}$, Ben Oppenheimer$^2$

\begin{abstract}
 
 We have measured the rest-frame $B,V,$ and $I$-band light
 curves of a high-redshift type Ia supernova (SN Ia), SN 1999Q
 ($z=0.46$), using {\it HST} and
 ground-based near-infrared detectors.  
  A goal of this study is the measurement of the color excess,
 $E_{B-I}$, which is a sensitive
 indicator of interstellar or intergalactic dust which could affect recent cosmological
 measurements from high-redshift SNe Ia.  Our observations disfavor 
a 30\% opacity of SN Ia visual light by dust
as an alternative to an accelerating Universe.  This statement applies to both Galactic-type dust (rejected at the 3.4$\sigma$ confidence level) and greyer dust (grain size $> 0.1 \ \mu$m; rejected at the 2.3 to 2.6 $\sigma$ confidence level) as proposed by Aguirre (1999).  The rest-frame $I$-band light curve shows the secondary
maximum a month after $B$ maximum typical of nearby SNe Ia of normal luminosity, providing no
 indication of evolution as a function of redshift out to $z \approx 0.5$.  An expanded set of similar
 observations could improve the constraints on any contribution of extragalactic dust to the dimming of high-redshift SNe Ia.

\end{abstract}
subject headings:  supernovae: general$-$cosmology: observations
 
\vfill
\eject
 
\section{Introduction}

     Recent observations of high-redshift ($z > 0.3$) Type Ia
 supernovae (SNe Ia) 
provide the backbone of the body of evidence that we live in an accelerating
Universe whose content is dominated by vacuum energy (Riess et
al. 1998; Perlmutter et al. 1999). 
The observational evidence for an accelerating Universe is that
high-$z$ SNe Ia are $\sim$30\% dimmer than expected in an open
Universe (i.e., $\Omega_M$=0.3, $\Omega_\Lambda=0$).  The
two most likely sources to obscure distant SNe Ia and affect their
interpretation are dust and evolution.  

The tell-tale
signature of extinction by Galactic-type dust, reddening, has not been
detected in the amount required to provide
$A_V=0.3$ mag for  high-$z$ SNe Ia (Riess et al. 1998; Perlmutter et
al. 1999).  Yet the cosmological implications of the observed
faintness of high-$z$ SNe Ia are so exotic as to merit the consideration
of dust with more unusual properties.  A physical model of dust composed of
larger grains ($>$ 0.1 $\mu$m) has been posited by Aguirre (1999a,b) 
to provide a non-cosmological source of extinction with less
reddening.  An {\it interstellar} component of this so-called ``gray'' dust,
if neglected,
would add
too much dispersion to be consistent with the observed luminosities
(Riess et al. 1998).
However, Aguirre (1999a,b) has shown that a uniformly distributed component
of {\it intergalactic} gray dust with a mass density of
$\Omega_{dust}\approx 5\times10^{-5}$ could explain the faintness of high-$z$
SNe Ia without detectable reddening and without overproducing the far-infrared (far-IR) background.  Previous data do not rule out this
possibility.  Indeed, significant {\it interstellar} extinction in the
hosts of high-$z$ SNe Ia is still favored by some (Totani \& Kobayashi 1999).

Rest-frame evolution is the other potential pitfall in
using high-$z$ SNe Ia to measure the cosmological parameters.
The lack of a complete theoretical model of SNe
Ia including the identification of their progenitor systems 
makes it difficult to access the expected evolution 
between $z=0$ and 0.5 (Livio 1999; Umeda et al. 1999; H\"{o}flich,
Wheeler, \& Thielemann 1998).
An impressive degree of similarity has been observed between the
spectral and photometric properties of nearby and high-$z$ SNe Ia (Schmidt et
al. 1998; Perlmutter et al. 1998, 1999; Riess et al. 1998; Filippenko et al. 2000; but
see also Riess et al. 1999c; Drell, Loredo, \& Wassermann 1999).
However, it is not known what kind or degree of change in the observable
properties of SNe Ia would indicate a change in the expected peak
luminosity by 30\%.  For that reason it has been necessary to compare a wide
range of observable characteristics of nearby and high-$z$ SNe Ia to
search for a complement to a luminosity evolution.

Near-IR observations of high-$z$ SNe Ia can provide constraints on both
sources of cosmological contamination.  A physical model of gray
intergalactic dust, such as that proposed by Aguirre (1999a,b), still
induces some reddening of SN light which can be detected in the
wavelength range between optical and near-IR light.  In
addition, near-IR observations provide a view of the behavior of 
high-redshift SNe Ia in a window previously unexplored.  Specifically, normal, nearby SNe Ia exhibit a second infrared maximum about a month after the
initial peak.   We can increase our
confidence that high-$z$ SNe Ia have not evolved by observing
this second maximum; its absence would indicate a change in the physics SNe Ia across redshift with potentially important cosmological consequences.

We obtained ground-based $J$-band
and space-based optical observations of SN 1999Q ($z=0.46$) to
initiate a study of the systematic effects of dust and evolution on
high-$z$ SNe Ia.  In \S 2 we descibe our observations, in \S 3 their
analysis, and in \S 4 their interpretation.

\section{Observations}

   Our High-$z$ Supernova Search Team (HZT) has an ongoing program to
discover and monitor high-redshift SNe Ia (Schmidt et al. 1998).  SN 1999Q was discovered on Jan
18, 1999 using the CTIO 4-m Blanco Telescope with the Bernstein-Tyson Camera (http://www.astro.lsa.umich.edu/btc/btc.html) as 
part of a 3-night program to search for high-$z$ SNe Ia using well-established methods.  High signal-to-noise ratio spectra of SN 1999Q
obtained with the Keck-II telescope indicated that this was a typical SN Ia
at $z$=0.46 shortly before $B$-band maximum (Garnavich et al. 1999; Filippenko et al. 2000).
  Rest-frame $B$ and $V$-band photometry using custom filters was obtained for
SN 1999Q from observatories around the world.  The {\it Hubble Space Telescope} ({\it HST}) monitored the $B$ and $V$ light curves of SN 1999Q from $\sim$
10 to 35 days after $B$ maximum (rest-frame) using the WFPC2 and the
F675W and F814W filters in the course of 6
epochs.  Combined, these data provide excellent coverage of the
rest-frame $B$ and $V$ light curves from a few days before to
60 days after $B$ maximum (in the rest frame; Clocchiatti et al. 2000).

  In addition, the SN was observed in the near-IR ($J$-band)
for 5 epochs between 5 and 45 days after $B$
maximum (in the rest frame).  The first observation employed the European Southern Observatory's 3.5-m New Technology Telescope
equipped with the Son of Isaac (SOFI) infrared camera spectrograph (http://www.ls.eso.org); subsequent observations
used the Keck II Telescope equipped with the Near-Infrared Camera (NIRC; Matthews \& Soifer 1994).  Due to
the high sky brightness in the IR, many dithered, short images were obtained and
combined to avoid saturating the detector.  Care was taken to maintain a
detected sky flux level of $\sim$10,000 counts, a regime where both SOFI and NIRC
exhibit less than 0.5\% non-linearity (http://www.ls.eso.org; http://www.keck.hawaii.edu).

  Using the procedure described by Garnavich et al. (1998), we
subtracted an empirical point-spread function (PSF) scaled to the brightness of the SN from
each {\it HST} observation.  A coadded image of total length 7200 seconds in
both F675W and F814W revealed no trace of host galaxy light to more
than 5 mag below the peak brightness of the supernova (i.e., $m_B,m_V >$ 27).   The host of SN 1999Q is likely
to be intrinsically faint or of very low surface brightness similar to the host of
SN 1997ck (Garnavich et al. 1998).  Due to the
negligible contribution of host galaxy light to the images, we have
made no correction for contamination to the measured supernova light. 
  Assuming that
the restframe $V-I$ color of the host galaxy is no redder than that of early-K-type dwarfs
($V-I$=1.0 for K0), this same practice is well justified for our
measurements of SN light in the $J$ band.  We conservatively adopt a systematic uncertainty of 0.03 mag in the SN
photometry (and 0.02 mag uncertainty in the colors) 
to account for any remaining bias.  The procedure described by Schmidt et al. (1998), Garnavich et
al. (1998), and Riess et al. (1998) was followed to calibrate the
measured magnitudes of SN 1999Q on the Johnson $B$ and $V$ passband
system. 

 Similar steps were performed to calibrate the observed $J$-band
magnitudes of SN 1999Q onto the rest-frame Cousins $I$-band system,
though a few exceptions are noted here.  On three photometric nights we
observed the secondary near-IR standards of Persson et al. (1998).  Because these
secondary standards are solar analogues (0.4 $<$ $B-V$ $<$ 0.8), one can transform these stars from the ``Persson system'' to that of
NIRC or SOFI by calculating the photometric difference of
spectrophotometry of the Sun between these systems.  We
found these differences to be quite small ($<$0.02 mag) and this
correction negligible.  In practice the true transmission curve in
$J$ is dictated by the natural opacity of atmospheric H$_2$O and
nightly variations are generally larger than differences between
different facility $J$-passbands.  For this reason we observed secondary standards in close temporal proximity to the SN field.  

Due to the inherent non-linearity of
airmass extinction corrections in $J$, the field of SN 1999Q was
observed at airmasses within 0.05 of the Persson et al. (1998) standards to avoid
the need for airmass corrections.  Assuming typical $J$-band
extinction of 0.1 mag per airmass (Krisciunas et al. 1987), errors of $\sim$
0.005 mag are introduced without explicit extinction corrections.

Cross-band $K$-corrections (Kim, Goobar, \& Perlmutter 1996) were calculated using spectrophotometry of
SN 1994D (which extend redward of 9100 \AA; Richmond et al. 1995)
to transform the $J$-band magnitudes of SN
1999Q to the Cousins $I$-band.  
At the redshift of SN 1999Q ($z=0.46$),
observed $J$-band light is an excellent match to rest-frame Cousins
$I$, and the $K$-correction was determined to be $-$0.93 $\pm0.02$ mag
with no apparent dependence on supernova phase or color.  The rest-frame $I$ photometry of SN 1999Q is given in Table 1.

Of fundamental importance is our ability to correctly transform 
the observed $J$-band photometry to restframe $I$-band.  Schmidt et al. (1998) discusses
the derivations of optical zeropoints from numerous spectrophotometric stars which
also have UBVRI photometry.  Applying the same spectrophotometry to cross-band $K$-corrections removes the dependence of the transformed photometry on the observed band's zeropoint.  Unfortunately, this type of data is not
available for the J-band. 
We have therefore calibrated our $J$-band data using Persson et al. standards, who are
fundamentally tied to the Elias (1982) standards and we adopt an appropriate $J$-band zeropoint uncertainty of $1-\sigma$=0.05 mag.

\section{Analysis}

\subsection{Second IR Maximum}

   A unique photometric signature of typical SNe Ia is a resurrection
of the luminosity at infrared wavelengths about a month after the initial maximum.  This feature is present (with the most exquisite photometry) in the $V$ band, grows into a ``shoulder'' in $R$, and increases to a second local
maximum in $I$ (Ford et
al. 1993; Suntzeff 1996; Hamuy et al. 1996a; Riess et al. 1999a).
  This second maximum is also readily apparent at
near-IR wavelengths ($J,H,$ and $K$; Elias et
al. 1981; Jha et al. 1999).  No other type of SN exhibits this
feature. 

 The secondary maximum is thought to result from the escape of
radiation from the core of the supernova at long wavelengths.
Resonant scattering from lines is the dominant source of opacity.  
At short wavelengths (i.e., $<$ 5000 \AA) line blanketing traps 
radiation; the resonance lines at longer wavelengths are fewer 
and further between providing escape routes for the trapped radiation (Spyromilio, Pinto, \& Eastman 1994). Wheeler et
al.(1998) argue that
this effect in itself would not explain the nonmonotic behaviour of the
J-band light curve. No model as yet fully explains the shape and timing
of the infrared light curves of SNe.

  The location and strength of the second $I$-band maximum is 
a diagnostic of the intrinsic luminosity of SNe Ia (Riess,
Press, \& Kirshner 1996; Hamuy et al. 1996b).  SNe Ia with typical peak luminosity (i.e., $M_V=-19.4$ mag) crest again in $I$ about 30 days after $B$ maximum.
Dimmer SNe Ia reach their second peak earlier.  For example, SN 1992bo was
$\sim$0.5 mag fainter than a typical SN Ia and reached its second peak in $I$ at $\sim$20 days after $B$ maximum (Maza et al. 1994).
For very subluminous SNe Ia this second maximum is completely absent,
merging into the phase of the initial decline (e.g., SN 1991bg;
Filippenko et al. 1992).   The physics detailing
the formation of this feature also indicates that its
magnitude and timing are sensitive to explosion parameters (e.g., ejecta composition) which determine the peak luminosity (Spyromilio, Pinto, \& Eastman 1994).

   In Figure 1 the relative rest-frame $I$-band magnitudes of SN 1999Q ($z=0.46$)
are compared to a luminosity sequence of nearby SNe Ia.  The
phase of the observations of SN 1999Q was determined by multicolor light-curve shape (MLCS; Riess et al. 1996; Riess et al. 1998) fits to the
$B$ and $V$ light curves and have
 an uncertainty ($1\sigma$) of less
than 2 days.  The observation times were also corrected for
$1+z$ time dilation (Leibundgut et al. 1996; Goldhaber et al. 1997; Riess et al. 1997).  Although the precision and sampling of the rest-frame $I$-band data are not
high, they are sufficient to indicate that this high-$z$ SN Ia
retains significant luminosity at $\sim$30 days after $B$ maximum,
consistent with the phase of the second $I$-band peak of typical SNe Ia and
inconsistent with either very subluminous or moderately subluminous
SNe Ia.  Using the $B$ and $V$ light-curve shapes of SN 1999Q as a
luminosity indicator, we find its distance modulus to be
$\mu_0$=42.67$\pm0.22$ mag, consistent with previous SNe Ia favoring a
cosmological constant (Riess et al. 1998).
If instead we consider the shape of the $I$-band light curve as an
 independent luminosity indicator, we find that this high-$z$ SN Ia (and presumably other high-$z$
SNe Ia) is not consistent with being subluminous by 0.5-0.6 mag as
needed to indicate a Universe closed by ordinary matter.  
More precise data will be needed to differentiate between
an open and $\Lambda$-dominated Universe solely on the basis of $I$-band light
curve shapes.

\subsection{IR Color Excess}

   To measure the $B-I$ colors of SN 1999Q we used the MLCS fits to
the $B$ light curve to determine the expected $B$ magnitudes at the
time of the IR observations.  Due to the exquisite {\it HST}
photometry in rest-frame $B$, this process adds little uncertainty
to the $B-I$ magnitudes.   

   The Milky Way (MW) dust maps from Schlegel, Finkbeiner, \& Davis (1998)
predict a reddening of $E_{B-V}$=0.021 mag in the direction of SN
1999Q.  We subtracted the expected Galactic reddening of the rest-frame
$B-I$ light (observed as $R-J$) of SN 1999Q, 0.037 mag, from the measured
colors.  Any remaining reddening results from extragalactic sources.  

   In Figure 2 the measured $B-I$ magnitudes of SN 1999Q are compared to a custom
$B-I$ curve predicted from the MLCS fits to the $B$ and $V$ light-curve shapes (Riess et al. 1996, 1998).  The smaller uncertainties shown here result from photon
statistics and were determined empirically (Schmidt et al. 1998).
  A significant, additional source of
uncertainty is the intrinsic dispersion of SNe Ia $B-I$ colors around
their custom MLCS model.  This intrinsic dispersion is determined
empirically by measuring the variance of 30 nearby SNe Ia around their
MLCS fits (Riess et al. 1996, 1998) and varies from 0.1 to 0.3 mag depending on the
SN Ia age.
Although the observed residuals from the model prediction are
correlated for time separations of less than 3 days, correlated errors
are insignificant for the larger differences in time between the observations of SN 1999Q.
The larger uncertainties shown in Figure 2 for the $B-I$
photometry of SN 1999Q include the
intrinsic uncertainties.  

   The measured $E_{B-I}$ for SN 1999Q is $-$0.09$\pm0.10$ mag.
The error includes the systematic uncertainties of the $K$-corrections and the
zeropoint of the $J$-band system, although the dominant sources of error are the photometry noise and the intrinsic dispersion in SN Ia $B-I$ colors.   
This
value is consistent with no reddening of this high-$z$ SN Ia.
If Galactic-type dust rather than a cosmological constant were the sole reason
that $z\approx0.5$ SNe Ia are 30\% fainter than expected for an open
Universe (i.e., $\Omega_M=0.3, \Omega_\Lambda=0.0$),
 then the $E_{B-I}$ of SN
1999Q should be 0.25 mag (Savage \& Mathis 1979).
  This alternative to an accelerating
Universe (see Totani \& Kobayashi 1999) is inconsistent with the data at the 99.9\% confidence level
(3.4$\sigma$).  The reddening required for the SNe Ia data to be
consistent with a Universe closed by matter is ruled out at the
$>$99.99\% (5.1$\sigma$) confidence level.
Despite the low precision of this data set, the
wavelength range of the $B-I$ colors results in the ability to rule out
extinction by Galactic-type dust from SN 1999Q alone with similar confidence as from the entire set of 
$B-V$ color data of Riess et al. (1998) and Perlmutter et al. (1999).  

   The reduced amount of reddening by ``gray'' dust grains (i.e., $>$ 0.1
$\mu$m) as proposed by
Aguirre (1999a,b) is more difficult to detect.  The amount of gray dust
needed to supplant the cosmological constant as the cause of the
dimming of high-$z$ SNe Ia would result in an $E_{B-I}$=0.17 or 0.14 mag
for a composition of graphite or graphite/silicate,
respectively (Aguirre 1999a,b).  These possibilities are moderately
inconsistent with the data at the 99.0\% (2.6$\sigma$) and  97.7\%
(2.3$\sigma$) confidence levels, respectively.  The reddening provided
by enough of such dust to change the cosmological forecast to
favor a Universe closed by matter is ruled out at the 99.97\%(3.7$\sigma$) and
99.90\%(3.3$\sigma$) confidence levels, respectively.  
The weakest constraint comes from assuming the smallest amount of
the grayest type of dust which is consistent at the 68\% (1$\sigma$)
confidence level with an open
Universe (i.e., $A_V$=0.2 mag).  This dust is inconsistent with
the data at the 94\% (1.9 $\sigma$) confidence level (although the
true inconsistency of this scenario is derived from the product of the two individual
likelihoods, i.e., 98\% or 2.3$\sigma$).  Although these results
disfavor the existence of the proposed levels of gray dust, more data
are needed to strengthen this important test.

   Because it is difficult to assess all sources of
uncertainty in our model for the SN Ia $B-I$ color evolution, we
also performed a Monte Carlo simulation of the measurement of
$E_{B-I}$ for SN 1999Q.  Using all nearby SNe Ia which are not
spectroscopically peculiar
(see Branch, Fisher, \& Nugent 1993) nor photometrically extreme ($0.9 <
\Delta m(B)_{15} < 1.6$; Phillips 1993) and whose $B-I$ colors were well observed, we
generated a standard, unreddened $B-I$ template
curve using individual reddening estimates from Phillips et al. (1999)
and Schlegel et al. (1998) for nearby
SNe Ia.  We then randomly selected five observations from a random
member of the sample and perturbed the observations
to match the photometric noise in the SN 1999Q observations.  From
10,000 such synthetic measurements we generated a distribution of
measured $E_{B-I}$ whose shape should match the probability density
function for the single $E_{B-I}$ measurement of SN 1999Q.  Compared
to the $B-I$ template curve, SN 1999Q has an $E_{B-I}$ = $-$0.12 mag.  The
distribution of synthetic $E_{B-I}$ values is asymmetric and implies an
uncertainty in the measurement for SN 1999Q  of $+1\sigma=0.11$ mag and $-1\sigma=0.17$ mag (including the systematic
uncertainties from $K$-corrections and the $J$-band zeropoint).  The
results are consistent with no extragalactic reddening and
inconsistent with Galactic and gray dust reddening at nearly identical
(though marginally higher) 
confidence levels as the MLCS fits.  The strength of this method is that it
samples real SN Ia data in the same manner as the observations of SN
1999Q and therefore incorporates the intrinsic and correlated
uncertainties in the $B-I$ colors of SNe Ia.

\section{Discussion}

   Two teams have independently concluded that the observed faintness
of high-$z$ SNe Ia indicates that the expansion of the Universe is
accelerating and that dark energy dominates the energy density of the
Universe (Riess et al. 1998; Perlmutter et al. 1999).  However, as a well-known
adage reminds us, ``extraordinary
claims require extraordinary evidence.''  Alternative explanations such as
evolution in supernova luminosities or dust are no more exotic than a cosmological
constant and must be rigorously tested.

\subsection{Dust}  

   A $\sim$30\% opacity of visual light by dust is the best quantified and therefore
most readily testable alternative to a cosmological constant (Aguirre
1999a,b; Totani \& Kobayashi 1999).
  Measurements of $B-V$
colors indicate that this quantity of Galactic-type dust is not obscuring
high-$z$ SNe Ia (Riess et al. 1998; Perlmutter et al. 1999) and the $B-I$ observations presented here bolster this evidence.   However, observations of neither SNe Ia nor other
astrophysical objects previously ruled out a similar opacity by
intergalactic gray dust (Aguirre 1999a,b).  The observations presented
here do disfavor a gray intergalactic medium providing this opacity,
but additional data are needed to strengthen these conclusions.
Indeed, a more precise measurement of $E_{B-I}$ or $E_{U-I}$ could
constrain either the total optical depth of dust in the intergalactic
medium or alternately push the minimum size of such grains into an unphysical domain (Aguirre 1999a,b).  It may even be possible to use such measurements to
constrain the contribution to the far-IR background by emission from the intergalactic
medium (Aguirre 1999a,b).  Measurements of gravitational lens systems have also been
used as a probe of the high-$z$ extinction law and disfavor
significant interstellar gray dust (Falco et al. 1999; McLeod 1999-except in molecular clouds).

\subsection{Evolution}

    To
date, our inability to formulate a complete theoretical description of
SNe Ia makes it impossible to either predict the degree of expected
luminosity evolution between $z=0$ and 0.5 or to identify an observation which would conclusively
determine whether the luminosity of SNe Ia are evolving (but see Hoeflich, Thielemann \& Wheeler 1998).  An empirical recourse is to 
compare all observable properties of nearby and high-$z$
SNe Ia with the assumption that if the luminosity of SNe Ia has
evolved by $\sim$30\% other altered characteristics of the explosion
would be visible as well.  The detection of such a change would cast
doubt on the reliability of the luminosity distances from
high-$z$ SNe Ia.  A continued failure to measure any difference between SN Ia near and far
would increase our confidence (though never prove) that evolution does not
contaminate the cosmological measurements from high-$z$ SNe Ia.

   Having clearly stated our approach, it is now appropriate to
 review the current status of the
ongoing efforts to determine if SNe Ia are evolving.

  Comparisons of high signal-to-noise ratio spectra of nearby and high-$z$ SNe Ia
 have revealed remarkable similarity
(Riess et al. 1998; Perlmutter et al. 1998,
1999; Filippenko et al. 2000).  Because the spectrum provides a
detailed record of the conditions of the supernova in the atmosphere (i.e.,
temperature, abundances, and ejecta velocities), spectral comparisons
are expected to be particularly meaningful probes of evolution.
Further, comparisons of time sequences of spectra reveal no apparent
differences as the photosphere recedes in mass (Filippenko et al. 2000), indicating that the
striking resemblence between distant and nearby SNe Ia is not merely
superficial, but endures at deeper layers.   
However, these comparisons still require the
rigor of a quantitative approach to determine
 whether or not the two samples are
statistically consistent.  The distributions of light-curve shapes at
high and low redshift are statistically consistent (Riess et al. 1998;
Perlmutter et al. 1999).  However, Drell et al. (1999) have noted that
different approaches to quantifying the shape of the light curves may
not be statistically consistent, so more attention needs to be focused on these light-curve shape comparisons.  

  The colors of pre-nebular supernovae should provide
 a useful probe of luminosity
evolution, indicating changes in the approximate temperature and hence
the thermal output of the explosion.  The $B-V$ colors of nearby and
 high-$z$ SNe Ia were found to be consistent by Perlmutter et
 al. (1999).  The same consistency was found here for the $B-I$
 colors.  However, neither this work nor the $B-V$ color measurements
 by Riess et al. (1998) can rule out the possibility that high-$z$ SNe
 Ia could be excessively blue (Falco et al. 1999); more data are
 needed to explore this possibility.

  The time interval between explosion and maximum light (i.e., the risetime)
is expected to be a useful probe of the ejecta opacity and the distribution of
$^{56}$Ni .  The initial comparison of the risetime of nearby (Riess et
al. 1999b) and high-redshift SNe Ia (Goldhaber 1998; Groom 1998) found an apparent inconsistency (Riess et al. 1999c).   Further analysis
of the SCP high-redshift data by Aldering, Nugent,
\& Knop (2000), however, concludes that the high-redshift risetime was somewhat larger and far more uncertain than found by Groom (1998) and that the remaining difference in the risetime could
be no more than a $\sim$2.0 $\sigma$ chance occurence.

   The weight of the evidence suggests no significant
evolution of the observed SNe Ia, but more observations
are needed to allay remaining reasonable doubts.  Perhaps the best
indication that SNe Ia provide reliable distances at high redshifts
comes from SNe Ia in nearby early-type and late-type galaxies.  These
galaxies span a larger range of metallicity, stellar age, and
interstellar environments than is expected to occur for galaxies back to $z=0.5$.  Yet after
correction for the light-curve-shape/luminosity relationship and
extinction, no significant Hubble diagram residuals are seen which
correlate with host galaxy morphology.  This suggests that
our distance estimates are insensitive to variations in the
supernova progenitor environment (Schmidt et al. 1998).  However,
the evidence remains circumstantial and does not rule out the
possibility that a characteristic of all progenitors of nearby SNe Ia
differs for high-$z$ SNe Ia.  

   Further observations, especially those in the near-IR bands, can better
constrain the potential contamination of the cosmological conclusions from SNe
Ia posed by dust and evolution.  Further, rest-frame $I$ band measurements of nearby SNe Ia show less dispersion in intrinsic luminosity and extinction making this an attractive band for future observations (Hamuy et al. 1996b).  
Measurements of SNe Ia at $z>1$ should even discriminate between the
effects of a cosmological constant and those of a monotonically increasing, but
unidentified systematic errors (Filippenko \& Riess 1999).  Continuing studies of high-$z$ SNe Ia should
ultimately provide
the extraordinary evidence required to accept (or refute) the accelerating Universe.

\bigskip 

We wish to thank Alex Athey and S. Elizabeth Turner for their help
in the supernova search at CTIO.  We have benefited from helpful
discussions with Anthony Aguirre, Stefano Casertano, and Ed Moran.
We thank the following for their observations or for attempts to obtain useful data A. Dey, W. Danchi, S. R. Kulkarni, \& P. Tuthill.
The work at U.C. Berkeley was supported by the Miller Institute for Basic
 Research
in Science, by NSF grant AST-9417213, and by grant GO-7505 from
the Space Telescope Science Institute, which is operated by the
Association of Universities for Research in Astronomy, Inc., under
NASA contract NAS5-26555.  Support for AC was provided by
the National Science Foundation through grant \#GF-1001-95 from AURA, Inc.,
under NSF cooperative agreement AST-8947990 and AST-9617036, and from
 Fundaci\'on
Antorchas Argentina under project A-13313.  This work was supported at
Harvard University through NSF grants AST-9221648, AST-9528899, and an NSF
Graduate Research Fellowship.  CS acknowledges the generous support of the
 Packard Foundation and
the Seaver Institute.  Based in part on observations collected at the
European Southern Observatory, Chile, under proposal 62.H-0324.

\vfill \eject
 
\centerline {\bf References}

\refitem Aguirre, A. 1999a, ApJ, 512, 19

\refitem Aguirre, A. 1999b, astro-ph/990439, accepted ApJ

\refitem Aldering, G., Nugent, P. E., \& Knop, R. 2000, submitted ApJ

\refitem Branch, D., Fisher, A., \& Nugent, P. 1993, AJ, 106, 2383

\refitem Clocchiatti, A., et al. 2000, in preparation

\refitem Drell, P. S., Loredo, T. J., \& Wasserman, I. 1999, astro-ph/9905027, accepted ApJ

\refitem Elias, J. H., Frogel, J. A., Hackwell, J. A., \& Persson, S. E. 1981, ApJ, 251, 13

\refitem Elias, J. H., 1982, AJ, 87, 1029

\refitem Falco, E. et al. 1999, ApJ, 523, 617

\refitem Filippenko, A. V. et al. 1992, AJ, 104, 1543

\refitem Filippenko, A. V. et al. 2000, in preparation

\refitem Filippenko, A. V., \& Riess, A. G. 1999, in {\it Type Ia Supernovae: Observations and Theory}. ed. J. Niemeyer and J. Truran (Cambridge: Cambridge Univ. Press), in press

\refitem Ford, C. et al. 1993, AJ, 106, 1101

\refitem Garnavich, P., et al. 1998, ApJ, 493, 53

\refitem Garnavich, P., et al. 1999, IAUC 7097

\refitem Goldhaber, G. 1998, B.A.A.S., 193, 4713

\refitem Goldhaber, G., et al. 1997, in {\it Thermonuclear Supernovae}, eds. P. Ruiz-Lapuente, R. Canal,  \& J. Isern (Dordrecht: Kluwer), p. 777

\refitem Groom, D. E. 1998, B.A.A.S., 193, 11102

\refitem Hamuy, M., et al. 1996a, AJ, 112, 2408

\refitem Hamuy, M., et al. 1996b, AJ, 112, 2438

\refitem H\"{o}flich, P., Wheeler, J. C., \& Thielemann, F. K. 1998,
ApJ, 495, 617

\refitem Jha, S., et al. 1999, ApJS, in press

\refitem Kim, A., Goobar, A., \& Perlmutter, S. 1996, PASP, 108, 190

\refitem Krisciunas, K., et al. 1987, PASP, 99, 887

\refitem Leibundgut, B., et al. 1996, ApJ, 466, L21

\refitem Livio, M. 1999, astro-ph/9903264

\refitem Matthews, K., \& Soifer, B. T. 1994, in {\it Infrared Astronomy with Arrays: the Next Generation}, ed. I. S. McLean (Dordrecht: Kluwer), p. 239 

\refitem Maza, J., Hamuy, M., Phillips, M., Suntzeff, N., \& Aviles, R. 1994, ApJ, 424, L107

\refitem Perlmutter, S., et al. 1998, Nature, 391, 51

\refitem Perlmutter, S., et al. 1999, ApJ, 517, 565

\refitem Persson, S. E., Murphy, D. C., Krzeminski, W., Roth, M., \& Rieke, M. J., 1998, AJ, 116, 2475

\refitem Phillips, M. M. 1993, ApJ, L105, 413

\refitem Phillips, M. M. et al. 1999, AJ, in press

\refitem Richmond, M. W. et al., 1995, AJ, 109, 2121

\refitem Riess, A. G., Press, W.H., \& Kirshner,  R.P. 1996, ApJ, 473,
88 

\refitem Riess, A. G., et al. 1997, AJ, 114, 722

\refitem Riess, A. G., et al. 1998, AJ, 116, 1009

\refitem Riess, A. G., et al. 1999b, astro-ph/9907037, accepted AJ

\refitem Riess, A. G., et al. 1999a, AJ, 117, 707

\refitem Riess, A. G., Filippenko, A. V., Li, W., \& Schmidt, B. P. 1999c, astro-ph/9907038, accepted AJ

\refitem Savage, B. D., \& Mathis, J. S. 1979, ARAA, 17, 73

\refitem Schlegel, D. J., Finkbeiner, D. P., \& Davis, M. 1998, ApJ, 500, 525

\refitem Schmidt, B. P., et al. 1998, ApJ, 507, 46

\refitem Spyromilio, J.,  Pinto, P., \& Eastman, R. 1994, MNRAS, 266, 17

\refitem Suntzeff, N. B. 1996, in {\it Supernovae and Supernovae Remnants}, ed. R. McCray \& Z. Wang (Cambridge: Cambridge Univ. Press), p. 41

\refitem Totani, T., \& Kobahashi, C. astro-ph/9910038, accepted ApJ

\refitem Umeda, H. et al. 1999, astro-ph/9906192

\refitem Wheeler, J. C., Hoeflich, P., Harkness, R. P., \& Spyromilio, J., 1998, ApJ, 496, 908

\vfill
\eject
 
{\bf FIGURE CAPTIONS:}
 
{\bf Fig 1.}-Relative rest-frame $I$-band light curve of a high-redshift SN Ia,
SN 1999Q ($z=0.46$), and a sequence of 3 nearby SNe Ia with different
light-curve shapes and peak luminosities.  The light curve of SN 1999Q
is consistent with that of typical nearby SNe Ia which reach their second maxima
at $\sim$30 days after $B$ maximum (e.g., SN
1995D; diamonds, Riess et al. 1999a).  The data are inconsistent with SNe Ia which are
subluminous at visual peak by $\sim$0.5 mag and reach the second
maximum at $\sim$20 days after $B$ maximum (e.g., SN
1992bo; asterisks, Maza et al. 1994).
  SN 1999Q bares no
resemblence to the rapid decline (without a second maximum) of very
subluminous SNe Ia (e.g., SN 1991bg; circles, Filippenko et al. 1992b).  

{\bf Fig 2.}-The color evolution, $B-I$, and the color excess, $E_{B-I}$, of a high-redshift SN Ia, SN
1999Q, compared to the custom MLCS template curve with no dust and enough dust (of either
Galactic-type or grayer) to nullify the cosmological constant.   The smaller error bars are from photometry noise; the larger error bars include all sources of uncertainty such as intrinsic dispersion of SN Ia $B-I$ color, $K$-corrections, and photometry zeropoints.  The
data for SN 1999Q are consistent with no reddening by dust, moderately
inconsistent with $A_V$=0.3 mag of gray dust (i.e., graphite dust with
minimum size $> 0.1 \ \mu$m; Aguirre 1999a,b) and $A_V$=0.3 mag of
Galactic-type dust (Savage \& Mathis 1979).  

\vfill
\eject

\begin{table}  
\begin{center}
\vspace{0.4cm}
\begin{tabular}{cccc}
\multicolumn{4}{c}{Table 1: Rest-Frame Photometry of SN 1999Q} \\
\hline
\hline
JD & Age$^a$ & $I$ & $B-I^{b}$ \\
(2451000+) & (day) & (mag) & (mag)  \\
\hline
204.2 &  +6.2    &       23.93(0.14)	& -0.62(0.15) \nl
216.4 &  +14.5  &        23.95(0.17) &   0.11(0.18)  \nl
239.3 &  +30.2   &       24.35(0.14)  &  1.30(0.15) \nl
243.3 &  +32.9   &       24.16(0.14) &   1.66(0.15) \nl
261.3 &  +45.3   &       24.59(0.19) &   1.65(0.20) \nl
\hline
\hline
\multicolumn{4}{l}{$^a$ Rest frame age relative to $B$ maximum} \\
\multicolumn{4}{l}{$^b$ $B$ mag fit from {\it HST} photometry} \\
\end{tabular}
\end{center}
\end{table}

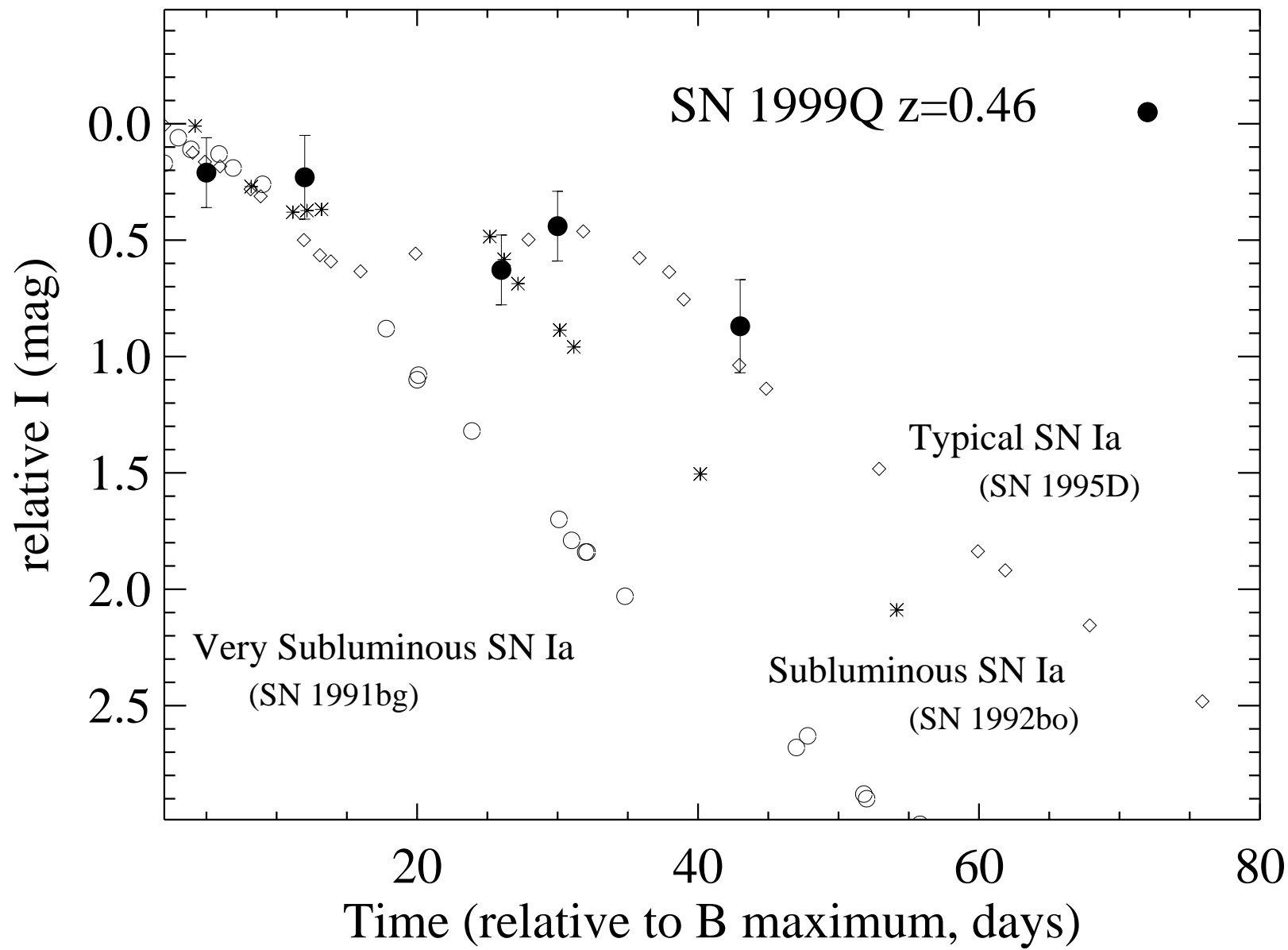
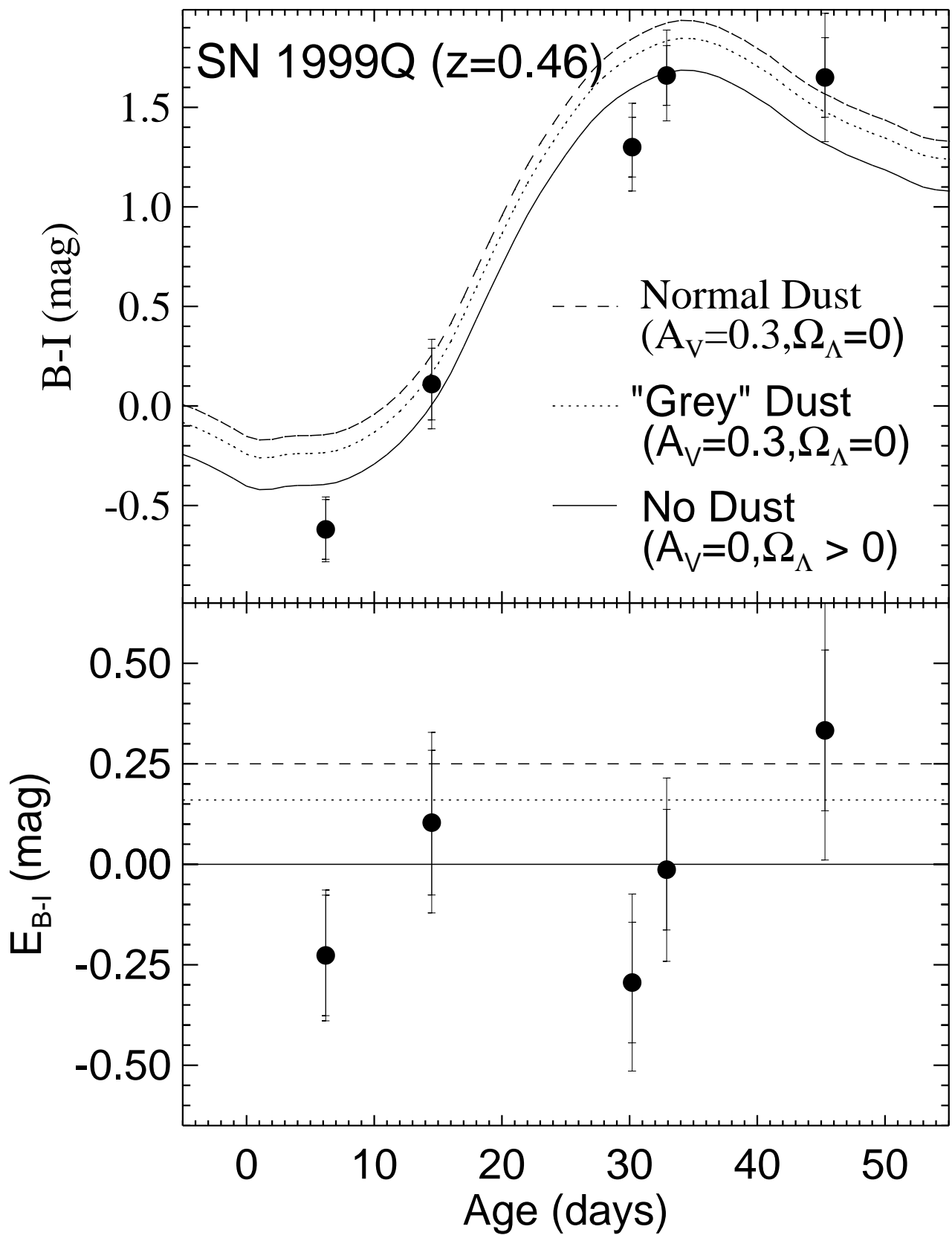
\end{document}